\documentclass[prl,aps,epsf,showpacs,twocolumn]{revtex4-1}
\usepackage{times}
\usepackage{graphicx}
\usepackage{float}
\usepackage{latexsym,amsmath,amssymb,bm,euscript}
\usepackage{color}
\usepackage{subfigure}
\usepackage{epstopdf}
\usepackage[colorlinks=true,linkcolor=blue,citecolor=blue,urlcolor=blue]{hyperref}
\usepackage{hyperref}
\usepackage[normalem]{ulem}
\usepackage[dvipsnames]{xcolor}
\usepackage{bm}
\usepackage{braket}
\usepackage{color,soul}
\usepackage{rotating,booktabs}
\usepackage{makecell}

\begin{document}
\title{Deep Learning-Enhanced Variational Monte Carlo Method for Quantum Many-Body Physics}

\author{Li Yang$^{1}$}
\email{lyliyang@google.com}
\author{Zhaoqi Leng$^{2}$}
\author{Guangyuan Yu$^{3}$}
\author{Ankit Patel$^{4,5}$}
\author{Wen-Jun Hu$^{1}$}
\email{nuaahwj@gmail.com}
\author{Han Pu$^{1}$}
\email{hpu@rice.edu}
\affiliation{
$^1$ Department of Physics and Astronomy \& Rice Center for Quantum Materials, Rice University, Houston, Texas 77005, USA \\
$^2$ Department of Physics, Princeton University, Princeton, New Jersey 08544, USA\\
$^3$ Department of Physics and Astronomy \& The Center for Theoretical Biological Physics, Rice University, Houston, Texas 77005, USA\\
$^4$ Department of Electrical and Computer Engineering, Rice University, Houston, Texas 77005, USA \\
$^5$ Department of Neuroscience, Baylor College of Medicine, Houston, Texas 77030, USA
}

\begin{abstract}
Artificial neural networks have been successfully incorporated into variational Monte Carlo method (VMC) to study quantum many-body systems. However, there have been few systematic studies of exploring quantum many-body physics using deep neural networks (DNNs), despite of the tremendous success enjoyed by DNNs in many other areas in recent years. One main challenge of implementing DNN in VMC is the inefficiency of optimizing such networks with large number of parameters. We introduce an importance sampling gradient optimization (ISGO) algorithm, which significantly improves the computational speed of training DNN in VMC. We design an efficient convolutional DNN architecture to compute the ground state of a one-dimensional (1D) SU($N$) spin chain. Our numerical results of the ground-state energies with up to 16 layers of DNN show excellent agreement with the Bethe-Ansatz exact solution. Furthermore, we also calculate the loop correlation function using the wave function obtained. Our work demonstrates the feasibility and advantages of applying DNNs to numerical quantum many-body calculations.
\end{abstract} 


\maketitle

{\it Introduction ---}
Over the past few years, artificial neural networks have been introduced to study quantum many-body systems~\cite{Carleo2017,Yusuke,Hiroki,caizi,Imada,Lixin,Neupert,Khatami}. In their seminal paper~\cite{Carleo2017}, Carleo and Troyer proposed to represent the quantum many-body states by a Restricted Boltzmann Machine (RBM), which contains one visible and one hidden layer. The many-body wave function is represented by the visible layer after integrating out the hidden layer. The parameters in the RBM are trained in the VMC method. Following this work, the RBM and a few other networks have been applied to study several quantum many-body systems with good accuracy~\cite{Carleo2017,Yusuke,Hiroki,caizi,Imada,Lixin,Neupert,Khatami,Torlai2018,Glasser2018,Kaubruegger2018}. So far, the networks that have been implemented in quantum physics studies are not deep and hence not powerful enough to represent more complicated many-body states. As a result, there has been no clear evidence that their performance far exceeds the more traditional state-of-the-art numerical algorithms sch as quantum Monte Carlo, density matrix reformalization group, or tensor networks, etc. To overcome this problem, deep neural networks (DNNs) have been suggested. Theoretical studies have shown that the DNNs can efficiently represent any tensor network states and most quantum many-body states, and possess distinct advantages over shallow networks~\cite{Duan,Clark2018,Levine2019}. In fact, the DNN-based deep learning has become the most successful model of many machine learning tasks and has dominated the field since 2012. The DNNs have been demonstrated to have comparable or superior performance in various tasks when compared to human experts, such as playing Atari games~\cite{Mnih2015},  Go~\cite{Silver2016,Silver2017}, and manipulating robots~\cite{Gu2016,Levine2016}, etc, and have led to rapid advances in artificial intelligence.  

Despite of great interest, there has been relatively few works in applying DNNs to quantum many-body computations \cite{Shashua, Carleo2019}. This perhaps is due to the fact that applying DNNs to represent quantum many-body states faces two main challenges: inefficient optimization and insufficient information for the proper choice of DNN architectures. The former arises because a DNN typically contains a large number of parameters to train, while a proper choice of the architecture often requires physical insights about the nature of the quantum systems. 

In this Letter, we propose an efficient convolutional DNN architecture to represent the ground state of quantum many-body systems. Most of the quantum systems consist of particles interacting with each other through finite range. Such a local interacting character can be ideally captured by the convolutional neural network (CNN). We have developed an importance sampling gradient optimization (ISGO) algorithm within VMC method, which significantly improves the optimization speed and hence enables us to utilize DNN architectures. Our method can take advantage of the automatic differentiation, which automatically computes gradient update via backward-propagation algorithm ~\cite{hecht1992theory}. We show that our method can be parallelized and take full advantage of the acceleration provided by graphics processing units (GPUs). The ISGO method achieves at least one order of magnitude speed up when trained on GPUs~\cite{suppl}.

For benchmark purpose, we construct DNNs to represent the ground-state wave function of the 1D SU($N$) spin chain, which has an exact solution under the Bethe ansatz. We systematically test different DNN architectures with ISGO for systems with different complexity. Our numerical results for the ground-state energies of 1D SU($N$) spin chain show excellent agreement with the exact solutions. Furthermore, we are able to compute correlation functions which are extremely difficult to obtain by Bethe Ansatz. The convolutional DNN architecture we constructed for this work can be readily generalized to represent the ground states of other quantum many-body systems. The ISGO method can also be used to accelerate the computation based on general VMC methods.

\begin{figure}[t]
\includegraphics[width=\columnwidth]{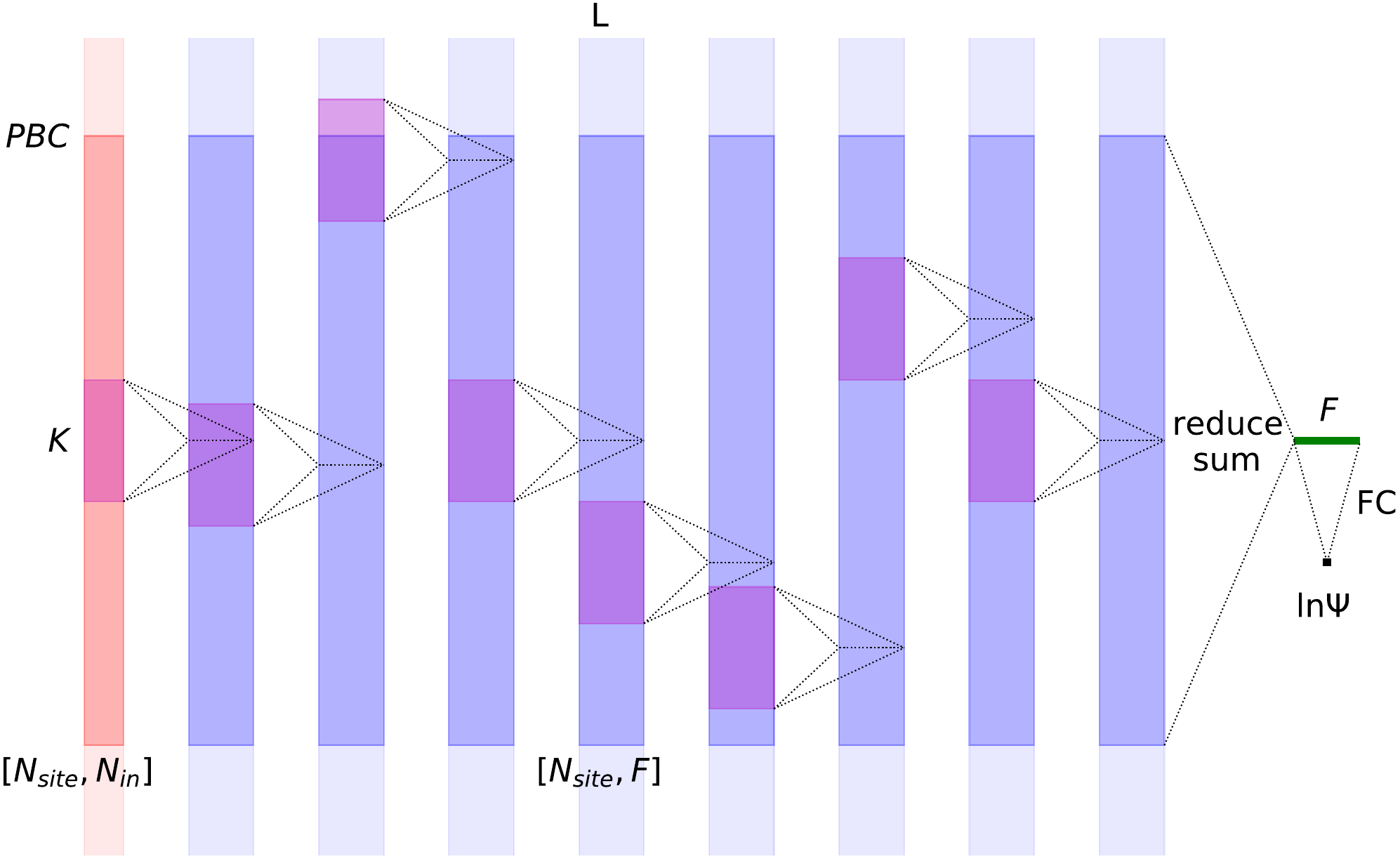}
\caption{The architecture of a convolutional DNN with $L=8$ hidden layers. The input state is encoded into a 2D tensor of shape [$N_\text{site}$, ${N}_\text{in}$], and fed into the input layer (represented by the leftmost pink rectangle). For value encoding $N_\text{in} = 1$ and for one-hot encoding $N_\text{in} = N$. The blue rectangles stand for the activation (feature maps) of the hidden layers. Convolution filters (the small pink rectangles) transform one hidden layer to the next one. The last hidden layer (on the right) is reduce summed and followed by a fully connected layer to give the $\text{ln}\Psi$ output.} \label{network}
\end{figure}

{\it Network Architectures ---} We consider a homogeneous 1D SU($N$) spin chain with $N_\text{site}$ spins, which is the simplest prototypical model with SU($N$) symmetry, governed by the Hamiltonian:
\begin{equation}\label{SUS}
H = \sum_{i=1}^{N_{\rm site}} P_{i, i + 1},
\end{equation}
where $P_{i, i + 1}$ is the spin exchange operator which exchanges two neighboring spins: $P_{i, i+1} |a_{i}, b_{i + 1}\rangle = |b_{i}, a_{i + 1}\rangle$. This model can describe the behavior of 1D strongly interacting quantum spinor gases~\cite{Deuretzbacher2014,Volosniev2014,Yang2015,Yang2016}, and has attracted significant attentions both experimentally and theoretically~\cite{Congjun,Honerkamp,gorshkov2010two,taie20126,pagano2014one,scazza2014observation,zhang2014spectroscopic,Hofrichter}. Here we will use the DNN to represent the ground-state wave function of this model.

A general state takes the form \[ |\Psi \rangle = \sum_{\{s_i\}} \Psi (s_1,s_2,...,s_{N_\text{site}})\, |s_1,s_2,...,s_{N_\text{site}}\rangle \,, \] where each $s_i$ represents one of the $N$ spin states for the SU($N$) model. The goal is to build a network that takes the basis state $|\{s_i\}\rangle$ as input and compute the ground state wave function $\Psi(\{s_i\})$ such that the energy functional $\langle\Psi|H|\Psi \rangle/\langle \Psi|\Psi\rangle$ is minimized. The first step is to encode the input basis state into a 2D tensor $S_{j,\beta}$, where the first and the second indices $j$ and $\beta$ represent the spatial site and the local spin state, respectively. In this work, we consider two kinds of state encodings: value encoding which encodes each spin state into a number and one-hot encoding which encodes the spin state into a one-hot boolean vector. The tensor $S$ is fed into the DNN as an input. The output of the first hidden layer, which follows immediately after the input layer, is given by:
\begin{equation}
\label{firstHidden}
A_{i,f'}^{[1]}=\sigma(\sum_{k=1}^{K}\sum_{f=1}^{N_\text{in}}W_{k, f, f'}^{[1]}S_{i+k,f}+b_{f'}^{[1]})\,,
\end{equation}
where $A^{[1]}$ is the activation (or feature map) of the first hidden layer, $\sigma(x) = \text{max}(x, 0)$ is the Rectified Linear Units (ReLU) activation function, which has been demonstrated to outperform traditional sigmoid activation function for DNNs~\cite{glorot2011deep}, $W^{[1]}$ is a 3D tensor of shape ($K$, $N_\text{in}$, $F$), where $K$ is the convolution kernel size, $F$ the number of channels of the hidden layer, and $N_\text{in}$ the number of the channels of the input layer which is 1 for value encoding and $N$ for one-hot encoding. In this work, we use the same $K$ and $F$, which determines the width of the network, for every hidden layer. $b^{[1]}$ is a bias vector of size $F$. The output from the remaining hidden layers are given by:
\begin{equation}
\label{hiddens}
A_{i,f'}^{[l]}=\sigma(\sum_{k=1}^{K}\sum_{f=1}^{F}W_{k, f ,f'}^{[l]}A_{i+k,f}^{[l-1]}+b_{f'}^{[l]})\,,\;\;l =2,3,...,L\,,
\end{equation}
where $L$ is the total number of hidden layers that determines the depth of the network, $W^{[l]}$ a 3D tensor of shape $[K, F, F]$, and $b^{[l]}$ a bias vector of size $F$. After the last hidden layer, its output is summed along the spatial dimension, and followed by a single fully connected layer to give the final output of the network
\begin{equation}
\label{output}
\text{ln}\Psi(S)=\sum_{f=1}^{F}a_{f}\sum_{i=1}^{N_{\text{site}}}A_{i,f}^{[L]}\,,
\end{equation}
where $a$ is a weight vector of size $F$. The full structure of the network is illustrated in Fig.~\ref{network}. Each magenta rectangular object corresponds to a convolutional filter. We use periodic padding for each convolutional layer to enforce periodic boundary condition. This network is fully convolutional~\cite{Long2014}, which means the network architecture is compatible with different system sizes, and we can easily do transfer learning. Here, $W^{[l]}$, $b^{[l]}$, and $a$ are the network parameters that need to be optimized. The total number of parameter is roughly $KF^2L$.

\begin{figure}[t]
\includegraphics[width=\columnwidth]{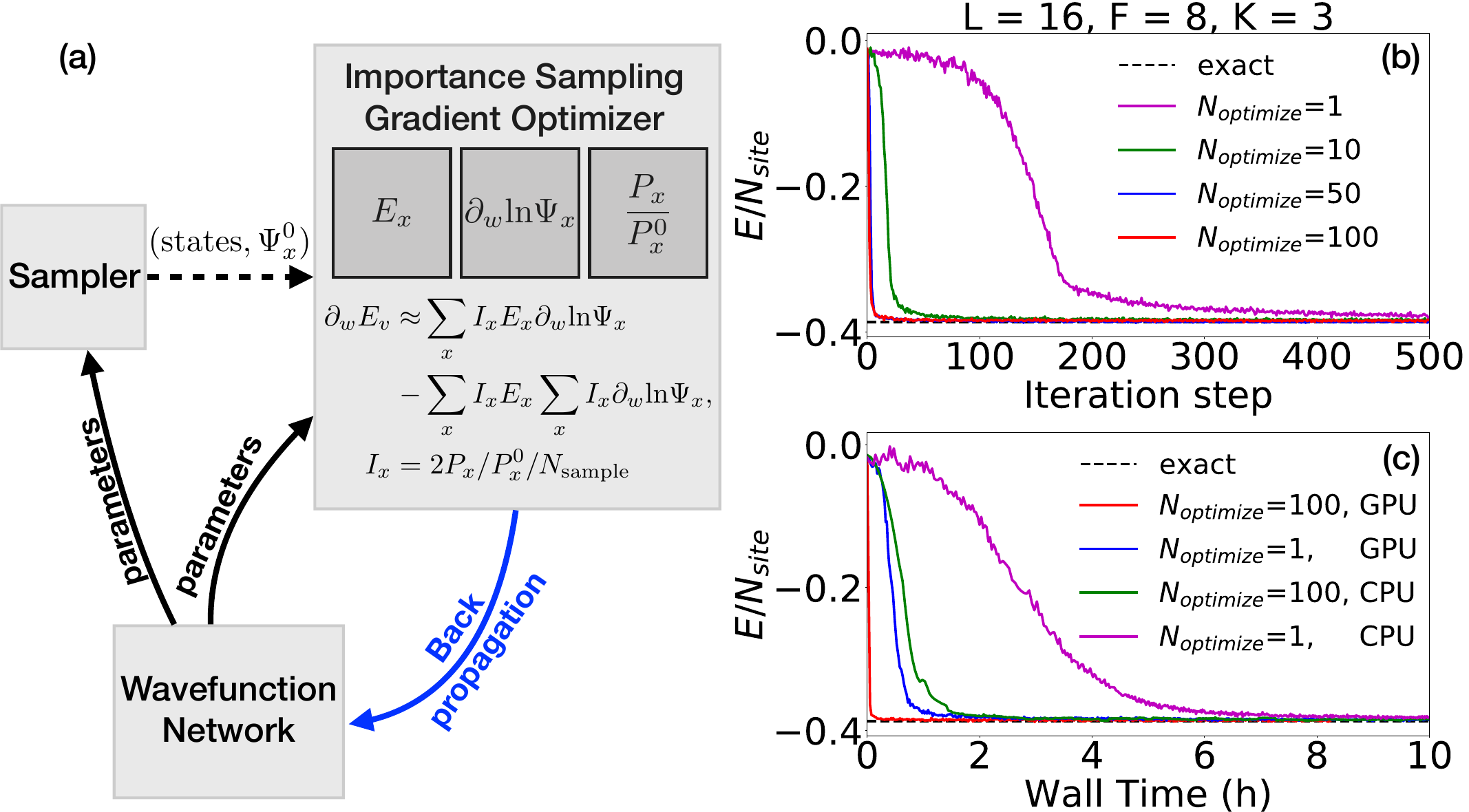}
\caption{(a) The flow chart of the ISGO algorithm within the VMC method. A network is used to represent the ground state wave function. In every iteration step, the sampler generates $N_\text{sample}$ samples following distribution $P^0_x \propto \Psi^{*0}_x\Psi^0_x$. Then the importance sampling optimizer updates the network parameters through back propagation in a loop for $N_\text{optimize}$ times. $N_\text{optimize}=1$ corresponds to the conventional gradient optimization method. The whole process is iterated until convergence. The sampler and the optimizer share the same wave function. We compare the training curves for a $N_{\rm site}=60$ SU(2) spin chain using a $ (L, F, K)=(16, 8, 3)$ CNN with one-hot encoding: (b) Variational energy versus iteration steps. (c) Variational energy versus wall time on GPU/CPU. The initial learning rate for Adam is $10^{-4}$.
}\label{offPolicy}
\end{figure}

{\it Importance Sampling Gradient Optimization ---} Before introducing the ISGO method, we first revisit the conventional gradient optimization method in VMC. The wave function $\Psi(\{w\})$ is encoded by the set of network parameters $w \in  \{W^{[l]}, b^{[l]}, a\}$. In every iteration step, $N_{\text{sample}}$ quantum states following distribution $P^0_x \propto |\Psi^0_x|^2$ are sampled using a Markov chain. Here $\Psi^0$ is the input wave function from the previous step, and $x$ indexes the sampled states. The variational energy functional $E_v(\{w\})= \frac{\bra{\Psi^0(\{w\})} H \ket{\Psi^0(\{w\})}}{\braket{\Psi^0(\{w\}) | \Psi^0(\{w\})}}$ is then computed. To minimize $E_v(\{w\})$, the network parameters are updated as $w \leftarrow w - \alpha \partial_w E_v$, where the ``learning rate" $\alpha$ is a small parameter~\cite{optimization}. In our work, we use the Adam \cite{Kingma2014}, a variant of the Stochastic Gradient Descent algorithm. Here, $\partial_w E_v$ is approximated by the variational wave function $\Psi^0$ with the $N_{\text{sample}}$ samples
\begin{equation}
\label{derivative}
\begin{split}
\partial_{w}E_v \approx \sum_x I^0 E^0_x\partial_w\text{ln}\Psi^0_x
- \sum_x I^0 E^0_x\sum_x I^0 \partial_w\text{ln}\Psi^0_x,
\end{split}
\end{equation}
where $E^0_x = \sum_{x'}H_{x,x'}\Psi^0_{x'}/\Psi^0_x$ is the local energy under $\Psi^0$ and $I^0 = 2 / N_\text{sample}$. After the parameters $w$ are updated, the wave function changes from $\Psi^0$ to $\Psi$ which serves as the input for the next iteration step, where a new set of states are sampled based on $\Psi$ and the previously sampled states based on $\Psi^0$ are discarded. 

Inspired by the off-policy policy gradient method in Reinforcement Learning (RL)~\cite{Meuleau2001,Sutton2018}, we develop an efficient importance sampling gradient optimization (ISGO) method that utilizes the mismatched samples, as shown in Fig.~\ref{offPolicy}(a). The key is to renormalize the distribution of those mismatched samples to $|\Psi_x|^2$ by multiplying the local energies and derivatives in Eq.~(\ref{derivative}) with importance sampling factors:
\begin{equation} 
\label{importanceSampling}
\begin{split}
\partial_{w}E_v \approx \sum_x I_x E_x\partial_w\text{ln}\Psi_x
- \sum_x I_x E_x\sum_x I_x \partial_w\text{ln}\Psi_x,
\end{split}
\end{equation}
where $E_x$ is the local energy under the last updated wave function $\Psi$, and $\frac{I_x}{I^0} = \frac{P_x}{P^0_x} = {\cal C}\frac{|\Psi_{x}|^2}{|\Psi_{x}^0|^2}$ with ${\cal C}$ the normalization factor which can also be approximated using $\sum_x I_x / I^0 = 1$. The key difference, in comparison to the conventional method, is that, within each iteration step, the network parameters $w$ (and hence the wave functions) are updated multiple times. This enables us to use the $N_{\rm sample}$ sampled states much more efficiently. Furthermore, the update procedure can be efficiently parallelized and run on GPUs.

We plot the variational energies versus iteration step and wall time for a 60 sites SU(2) chain with a 16 layers CNN in Fig.~\ref{offPolicy}(b) and (c)~\cite{suppl}. As can be seen, the ISGO method converges with much fewer samples and much faster on GPU than the conventional method. We also implement the ISGO method using Tensorflow with auto differentiation \cite{stop_gradient}, which allows us to try new network architectures much more easily. Our code for both RBM and CNN can be found in Ref.~\cite{code}. We emphasize that, although we choose a particular gradient optimization algorithm Adam in our work, the concept of ISGO is general and can be implemented with any other optimization methods.

\begin{figure}[t]
\includegraphics[width=\columnwidth]{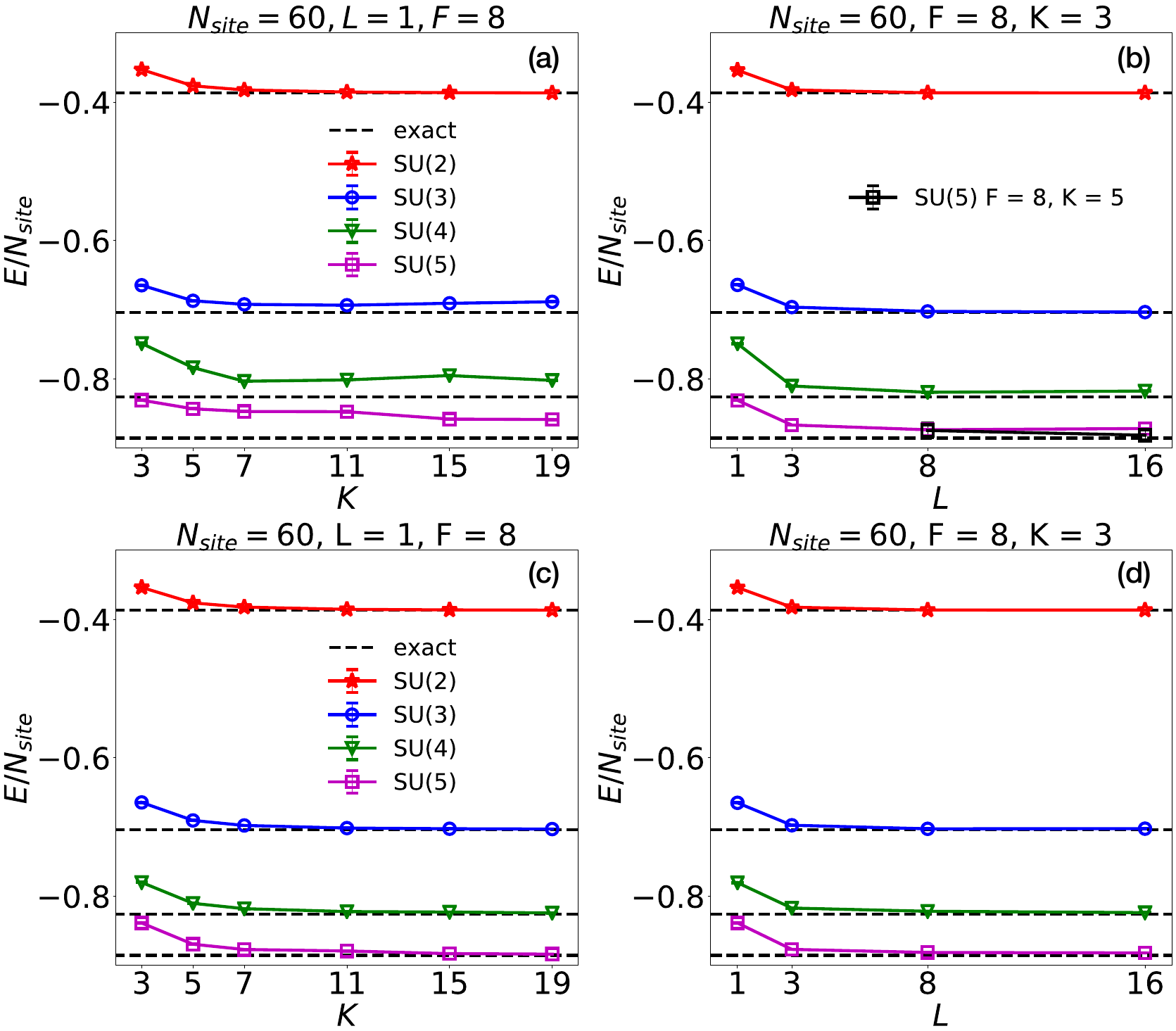}
\caption{The ground-state energy for $N_\text{site}=60$ SU($N$) spin chains ($N = 2, 3, 4, 5$) using CNN with value encoding (a)(b) and one-hot encoding (c)(d). (a) and (c) are for one layer $\text{L}=1$ with fixed channel number $F=8$ and different kernel size $K$. (b) and (d) are for fixed channel number and kernel size ($F=8$, $K=3$) but different number of layers $L$. In (b), the black squares are for $F=8$ and $K=5$. The horizontal black dashed lines are exact results from Bethe Ansatz.}\label{energyResults}
\end{figure}

{\it Numerical Results for 1D $SU(N)$ Spin Chain ---}
We test our DNN on the Sutherland model, the 1D homogeneous SU($N$) spin chain governed by Hamiltonian (\ref{SUS}). We pick this model for two main reasons. First, the ground state energy of this model can be exactly solved by Bethe Ansatz~\cite{Sutherland1975}, which allows us to benchmark our results~\cite{suppl}. Second, the number of spin states $N$ controls the complexity of the system, which allows us to systematically study the efficiency and accuracy of the DNN as the complexity of model grows. Numerical details can be found in \cite{suppl}.


Figure~\ref{energyResults} shows our main results of the ground-state energies for various $N$ on an ${N}_\text{site}=60$ chain with a DNN with varying depth (i.e., number of layers $L$) and width (i.e., kernel size $K$). We tested two encoding methods for the input state. Fig.~\ref{energyResults}(a,b) correspond to the value encoding, while (c,d) correspond to the one-hot encoding. The value encoding imposes ordinality, i.e., different spin states are encoded into a number with the average value to be zero. For one-hot encoding, different spin states are encoded into a vector orthogonal to each other, and thus are not ordinal. For example, for an $N=3$ system, the three spin states are encoded into values of $-1/2$, 0, $1/2$ in value encoding, and into vectors $(1, 0, 0)$, $(0, 1, 0)$, $(0, 0, 1)$ in one-hot encoding. The one-hot encoding requires more computational resources (in terms of both memory and computational time) than the value encoding, and scales lineally with respect to $N$. However, in general it yields better accuracy than the value encoding method. This could be due to the fact that one-hot encoding encodes each spin state into a vector which effectively enlarges the dimension of the parameter space. Optimization in such an artificially enlarged space helps to prevent the system stuck in metastable states \cite{Pickard2019}.
 
Figures~\ref{energyResults}(a,c) display the results from a single-layer network with varying width. As one can see, increasing the kernel size $K$ helps bring the ground-state energy closer to the exact solution represented by horizontal dashed lines, which indicates that, for such a shallow network, large kernel size is necessary for capturing long-range effects mediated by nearest-neighbor interactions. Here, one-hot encoding performs significantly better than value encoding especially for SU($N>2$), where, no matter how large the kernel size is, the energies computed via value encoding do not converge to exact solutions. In Fig.~\ref{energyResults} (b, d), we fix the width of the network, but vary its depth by adjusting the number of layers $L$. Even for a relatively small kernel size $K=3$, increasing $L$ helps to bring the computed ground-state energy closer to the exact result. Therefore, a DNN can capture long-range effect even with a small kernel size. For the SU(5) model (the largest $N$ we used in the calculation), the energy does not converge to exact solution using value encoding with kernel size 3. Simply by increasing the kernel size to 5, we can reduce the computed ground-state energy and match it with the exact solution as the black squares shown in Fig.~\ref{energyResults}(b). We vary the number of channels $F$ and find that the energy results are not sensitive to $F$. More details can be found in the Supplemental Materials~\cite{suppl}.
 
The Bethe Ansatz method can yield energy spectrum and, in principle, the many-body wave function for exactly solvable models such as the one considered here. However, due to the complexity of a general many-body wave function, it remains a tremendous challenge to compute other useful quantities such as the correlation functions. Often, advanced numerical techniques are needed for such tasks~\cite{mila2015, mila2018}. To further demonstrate the power of DNN, here we show our results for the loop correlation function
\begin{equation}
S_{m,n}=(-1)^{m - n}\,\langle(m\cdots n) \rangle\,,\label{loopCorr}
\end{equation}
where, the expectation value is taken with respect to the ground state, and $(m\cdots n)$ is the loop permutation operator that permutes the spatial indices in the wave function by $m\rightarrow m+1,m+1\rightarrow m+2,...,n-1\rightarrow n,n\rightarrow m$. Physically, this operator puts the spin in the original $n^{\rm th}$ position to the $m^{\rm th}$ position and correspondingly move the spins at the original $i^{\rm th}$ (with $i=m, ..., n-1$) positions to their neighboring position on the right. The loop correlation function appears in the definition of the one-body density matrix of 1D strongly interacting quantum spinor gases whose ground state can be represented by a strong coupling ansatz wave function~\cite{Levinsen2015,Yang2015,Yang2016,Yang2017, Deuretzbacher2014} due to the fact that such wave functions must obey the permutation symmetry rule originated from quantum indistinguishability. 

\begin{figure}[t]
\includegraphics[width=\columnwidth]{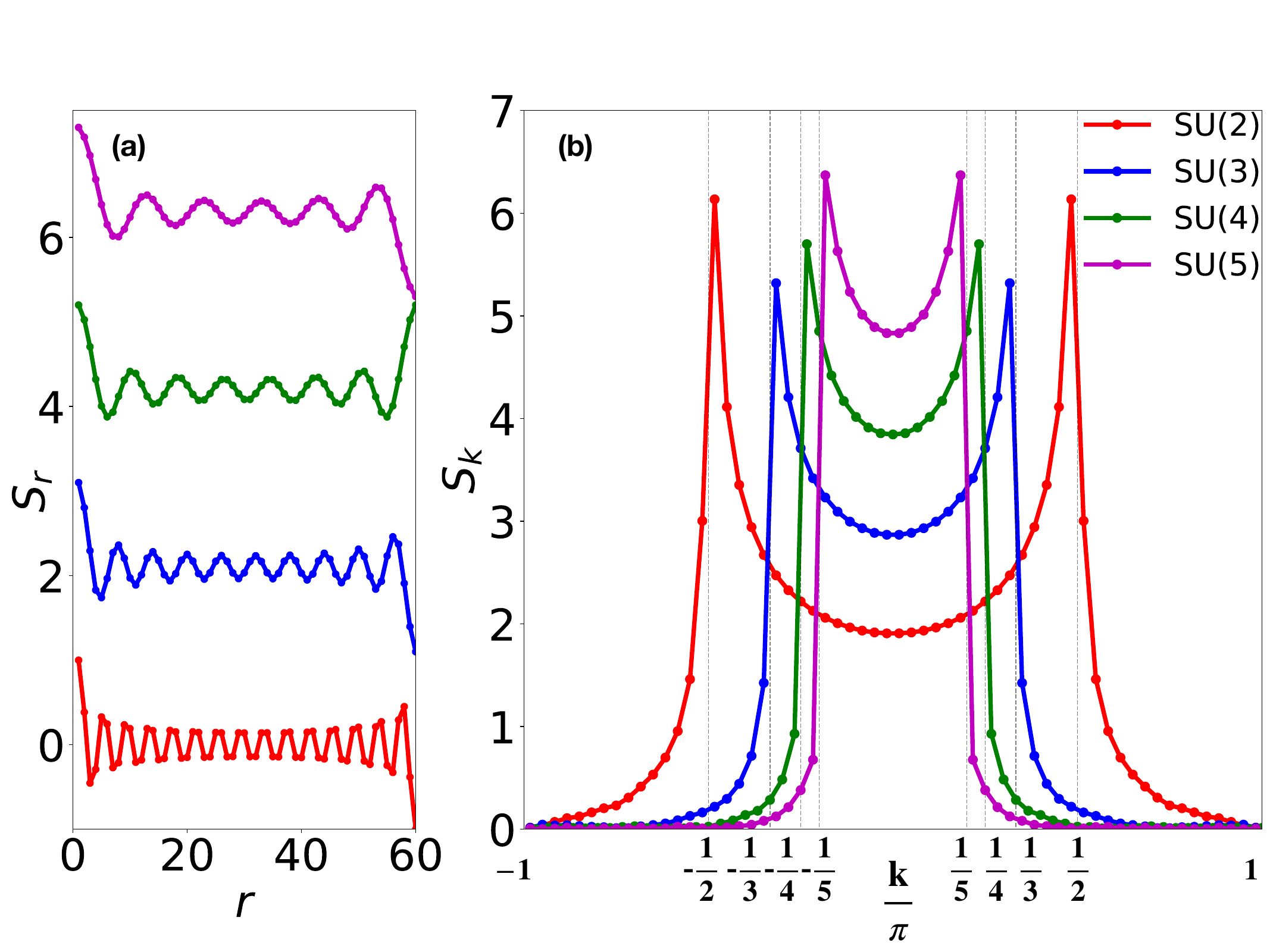}
\caption{(a) Real space loop correlation functions $S_r$ for the SU($N$) spin chain with $N=5, 4, 3, 2$ from top to bottom. (b) The Fourier transform of $S_r$:  $S_k = |\sum_r S_r e^{ikr}|$ with peaks at $k=\pm \pi/N$.}\label{loop}
\end{figure}

For the homogeneous system we considered here, $S_{m,n} = S_r$ with $r \equiv n-m$. We plot $S_r$ for an SU($N$) spin chain with $N_\text{site} = 60$ spins in Fig.~\ref{loop}(a), and its discrete Fourier transform $S_k = |\sum_r S_r e^{ikr}|$ in Fig.~\ref{loop}(b). $S_r$ and $S_k$ characterize the correlation in the real and the momentum space, respectively. By taking the Jordan-Wigner transformation, an SU($N$) spin chain with each spin component having $N_\text{site}/N$ spins can be mapped to a nearest-neighbor interacting $N$-component fermionic system with each component having $N_\text{site}/N$ fermions~\cite{Ogata1990,Yang2017}. The peaks of $S_k$ at $k=\pm \pi/N$, that can be clearly seen in Fig.~\ref{loop}(b), correspond to the Fermi points of those fermions. 
These peaks lead to the singularities of momentum distribution of strongly interacting spinor Fermi gases at the same momentum point \cite{Ogata1990,Yang2017}.

{\it Conclusion and Outlook ---} We have constructed a DNN, combined with VMC, to study the ground state of the 1D SU($N$) spin chain. The key in our work is the development of the ISGO algorithm, which can be straightforwardly applied to any type of variational wave functions, for the optimization procedure. This algorithm allows us to efficiently train the network, and is particularly suitable for training DNNs which typically contains a large number of parameters. Note that the VMC with the ISGO algorithm may be interpreted as an RL process if we identify the Markov-chain state trajectories in the former as the state transitions/policies in the latter. We tested the network to solve the 1D SU($N$) spin chain model and systematically investigated the performance of the network by varying its depth and width. We have found that, when using value state encoding, as the complexity of the model increases by increasing $N$, it is not sufficient just to increase the width (i.e., kernel size) of the network, one needs to add more depth to capture the long-range correlation of the quantum state. We only show numerical results computed by the DNN up to 16 layers. We do not observe any significant benefit by using much deeper networks up to 100 layers on this model. This could be due to a potential problem of vanishing gradients in very deep networks (see Ref.~\cite{srivastava2015training, He2016} and references therein), which may be alleviated via using other network architectures such as ResNet \cite{He2016}, which we leave for future studies. Finally, we note that another key finding from our work, which has not been discussed in previous works, is the importance of input states encoding. We find that one-hot encoding, although requires more computational resource, in general leads to much more accurate results than value encoding. 

In conclusion, our study clearly demonstrates that it is feasible to use DNNs to represent quantum many-body wave functions and to significantly enhance the efficiency of numerical quantum many-body computation. Applying machine learning techniques to quantum many-body physics is still a young and emerging field with many open questions. We believe that such investigation will not only benefit quantum physics, but may also help us to gain deeper insights into the neural networks.


\begin{acknowledgments}
{\it Acknowledgement ---} 
H.P. acknowledges support from NSF and the Robert A. Welch Foundation (Grant No. C-1669), W.-J.H. is supported by the DOE Award No.\ DE-SC0018197 and the Robert A. Welch Foundation (Grant No.\ C-1411). The majority of the computational calculations have been performed on the K80 GPUs with 11GB memory of Google Colaboratory. The work was finished when L.Y. was at Rice, who is now at Google.
\end{acknowledgments}

\bibliography{dnnsun}

\begin{thebibliography}{54}
\bibitem [{sto()}]{stop_gradient}%
  \BibitemOpen
  \href@noop {} {}\bibinfo {note} {A loss function ${\cal L} = \sum_xI_x
  E_x\text{ln}\Psi_x - \sum_xI_xE_x\sum_xI_x\text{ln}\Psi_x$ can be defined,
  and tf.stop\_gradient() can be applied on $I_x=P_x/P^0_x/N_\text{sample}$ and
  $E_x$ to keep the gradient only flow back through
  $\text{ln}\Psi_x$.}\BibitemShut {Stop}%
\end{thebibliography}


\begin{center}
\begin{large}
{\bf Supplemental Material}
\end{large}
\end{center}

\section{Bethe Ansatz Solution for Ground-State Energy of SU($N$) Spin Chain}
The 1D homogeneous SU($N$) spin-chain model (Sutherland model) can be solved using Bethe-Ansatz as reported in Ref.~\cite{Sutherland1975}. In this section, we provide details of how to numerically calculate the ground-state energy for a finite size SU($N$) spin-chain model with periodic boundary condition. Consider the case with all $N$ spin components equally populated with M spins. There are a total number of $N_\text{site} = MN$ spins. Following Eq.~(63)($\alpha$)($\beta$)($\zeta$) in Ref.~\cite{Sutherland1975}, the Bethe Ansatz equations consist of the following $(N - 1)$ set of coupled equations:
\begin{align}
\begin{split}
\label{BAE}
&N\theta(2\alpha)-\sum_{\alpha'}\theta(\alpha-\alpha')+\sum_{\beta}\theta(2\alpha-2\beta)+2\pi J_{\alpha}=0\,, \\
&\sum_{\alpha}\theta(2\beta-2\alpha)-\sum_{\beta'}\theta(\beta-\beta')+\sum_{\gamma}\theta(2\beta-2\gamma)+2\pi J_{\beta}=0\,, \\ 
&\vdots \\
&\sum_{\delta}\theta(2\zeta-2\delta)-\sum_{\zeta'}\theta(\zeta-\zeta')+2\pi J_{\zeta}=0
\end{split}
\end{align}
where $\theta(x)=-2\text{arctan}(x)$. $\alpha, \beta, \gamma, ...$ are the $(N - 1)$ sets of rapidities to be solved. The $j^{\rm th}$ set of equations (rapidities) has size $M_j = (N - j) M$, where $j = 1, 2, ..., N - 1$. For example, there are $(N-1)M$ different equations for $(N-1)M$ different $\alpha$'s, $(N-2)M$ different equations for $(N-2)M$ different $\beta$'s, and so on. So there are totally $(N-1)NM/2$ different equations. $J_{\alpha}$'s are integers or half odd integers and serve as quantum numbers \cite{Sutherland1975}. For $N= 2$, there is only a single set of equations with size $M$:
\begin{equation}
N\theta(2\alpha)-\sum_{\alpha'}\theta(\alpha-\alpha')+2\pi J_{\alpha}=0\,,
\end{equation}
which are just the Bethe Ansatz Equations for 1D Heisenberg model.

For every set of equations with size $M_{j}$, the quantum numbers $J$'s are concentrated around the origin. 
\begin{equation}
J_{j}=-\frac{M_{j}-1}{2},\,-\frac{M_{j}-1}{2}+1,\,...\,,\,\frac{M_{j}-1}{2}\,.
\end{equation}
And the ground state energy can be calculated by Eq.\,(56) in Ref.~\cite{Sutherland1975}:
\begin{equation}
E=N-2(N - 1)M-2\sum_{l=1}^{(N - 1)M}\text{cos}K_{l}\,.
\end{equation}
Using the the relation of momentum and rapidity: $K_{l}=-\theta(2\alpha_{l})$, we have:
\begin{equation}
\label{energy}
E = N-\sum_{l=1}^{(N-1)M}\frac{1}{1/4+\alpha_{l}^{2}}\,.
\end{equation}
Note that the rapidities in Eq.~(\ref{BAE}) can be numerically solved for up to hundreds of spins using standard zero-finding libraries such as SciPy. Therefore, the exact ground state energy can be efficiently computed via Eq.~(\ref{BAE}) and Eq.~(\ref{energy}). We provide the python code for solving the ground state energy for arbitrary $N$ and $N_\text{site}$ in \cite{code}.

\section{The Unitary Transformation}
In general, the ground-state wave function of the SU($N$) spin model Eq.~(\ref{SUS}) in the main text is not always positive. However, current DNNs only support positive wave functions. The reason is that at the final output, first $\text{ln}\Psi$ is calculated then an exponential is used to output $\Psi$, which is a positive value. In order to overcome this constraint, one needs to either add sign in the DNN by hand or do a unitary transformation for the Hamiltonian. We choose the latter by replacing the spin exchange operator $P$ in Eq.~(\ref{SUS}) of the main text by
\begin{equation}
\tilde{P}\left|\alpha,\beta\right\rangle =\begin{cases}
\left|\beta,\alpha\right\rangle  & \alpha=\beta\\
-\left|\beta,\alpha\right\rangle  & \alpha\ne\beta
\end{cases}
\end{equation}
The new Hamiltonian has exactly the same eigenspectrum, but its ground-state wave function is positive. 
In this paper, we focus on studying the new model with the spin exchange operator $\tilde{P}$ under periodic boundary conditions. And we also need to change the loop correlation operators in Eq.~(\ref{loopCorr}) of the main text by adding corresponding signs. 

\section{More Numerical Details and Results}
We typically use 1500 iteration steps, the initial learning rate for Adam is set to $10^{-3}$ for the first 100 steps and then changed to $10^{-4}$, $N_\text{sample}=20000$ except for the first 10 steps to avoid out-of-memory, and $N_\text{optimize}=100$ for the first 1000 steps and then set to 10. We average the network parameters over the last 100 steps when calculating physical quantities.

\begin{figure}[t]
\includegraphics[width=\columnwidth]{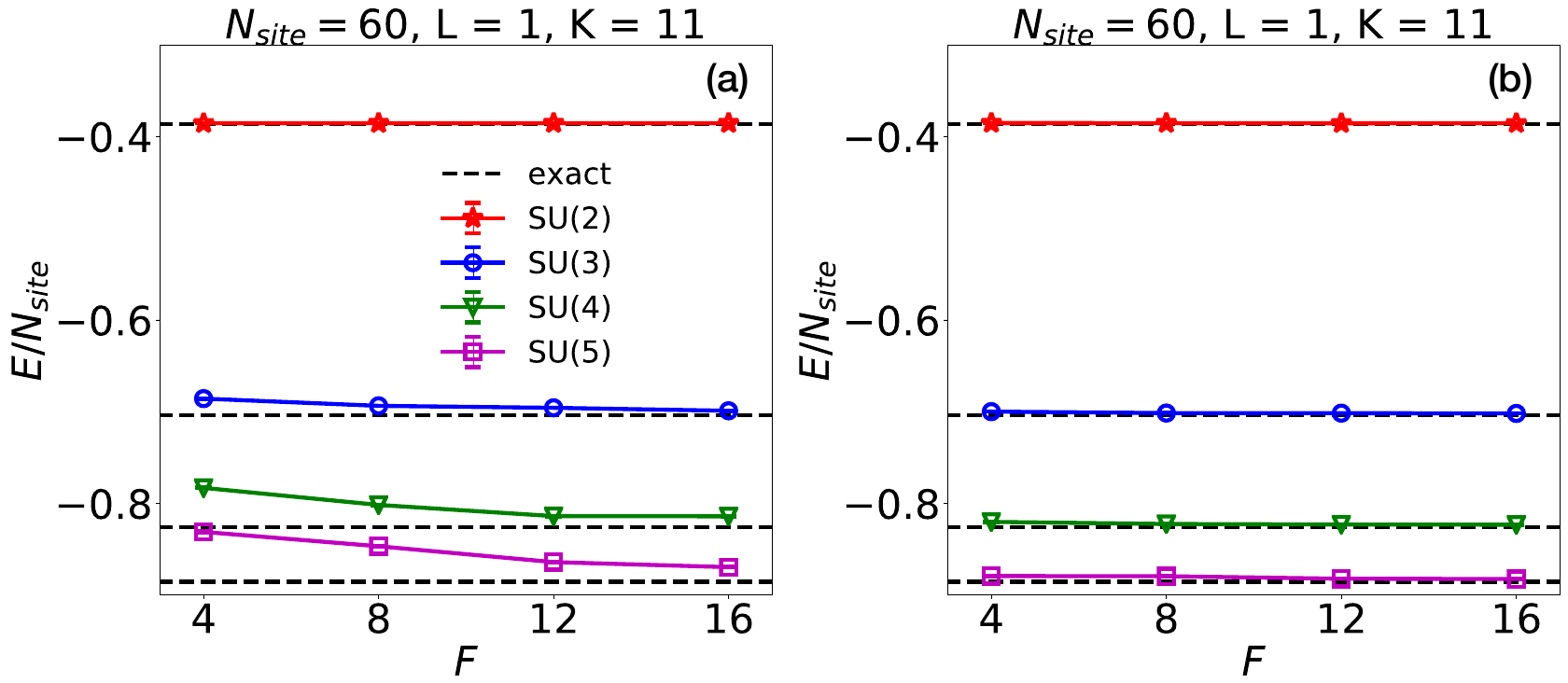}
\caption{The 1-hidden layer CNN ground state energies for 60 sites SU($N$) spin chains, where $N = 2, 3, 4, 5$. (a) is for state value encoding. (b) is for state one-hot encoding. The only variable parameter is the number of filters F. The black dashed lines are exact results from Bethe Ansatz.}\label{energyResultsF}
\end{figure}

Besides the kernel size $K$ and number of layers $L$, we have also studied the effects of changing the number of filters $F$. We fix the kernel size $K=11$ with one hidden layer, and the results are shown in Fig.~\ref{energyResultsF}. We found that for the value encoding in Fig.~\ref{energyResultsF}(a), the energies are not sensitive for SU(2) and SU(3) models. But for SU(4) and SU(5) models, the variational energies are still far away from the exact solutions. While for the one-hot encoding in Fig.~\ref{energyResultsF}(b), all energies are not sensitive to number of filters.


\begin{table}[h!]
\centering
\begin{tabular}{ |c|c|c|c|c|c| } 
\hline
\thead{$N$} & \thead{BA} & \thead{one-hot encoding} & \thead{error [\%]} & \thead{value encoding} & \thead{error [\%]} \\ 
\hline
2 & -0.3868 & -0.3865 & 0.06 & -0.3866 & 0.05 \\ 
3 & -0.7038 & -0.7027 & 0.16 & -0.6885 & 2.18 \\ 
4 & -0.8258 & -0.8237 & 0.25 & -0.8020 & 2.88 \\ 
5 & -0.8854 & -0.8827 & 0.32 & -0.8586 & 3.03 \\ 
\hline
\end{tabular}
\caption{Comparison of the variational energies of a one-hidden layer CNN for $N_\text{site}=60$ SU($N$) spin chains with exact Bethe Ansatz (BA) results. The kernel size is $K=19$. Both results from the one-hot and the value encoding are included. The incertitudes on the VMC data are smaller than $10^{-4}$.}\label{energies_compare}
\end{table}

In Table I, we list the ground state energies obtained from our numerical calculation with a one-layer CNN and compare them with the Bethe Ansatz (BA) exact results. One can clearly see that the one-hot encoding yields much more accurate results than the value encoding. The one-hot results are comparable to those given in Ref.~\cite{mila2015}.

\section{Comparison with Restricted Boltzmann Machines}
\begin{figure}[t]
\includegraphics[width=\columnwidth]{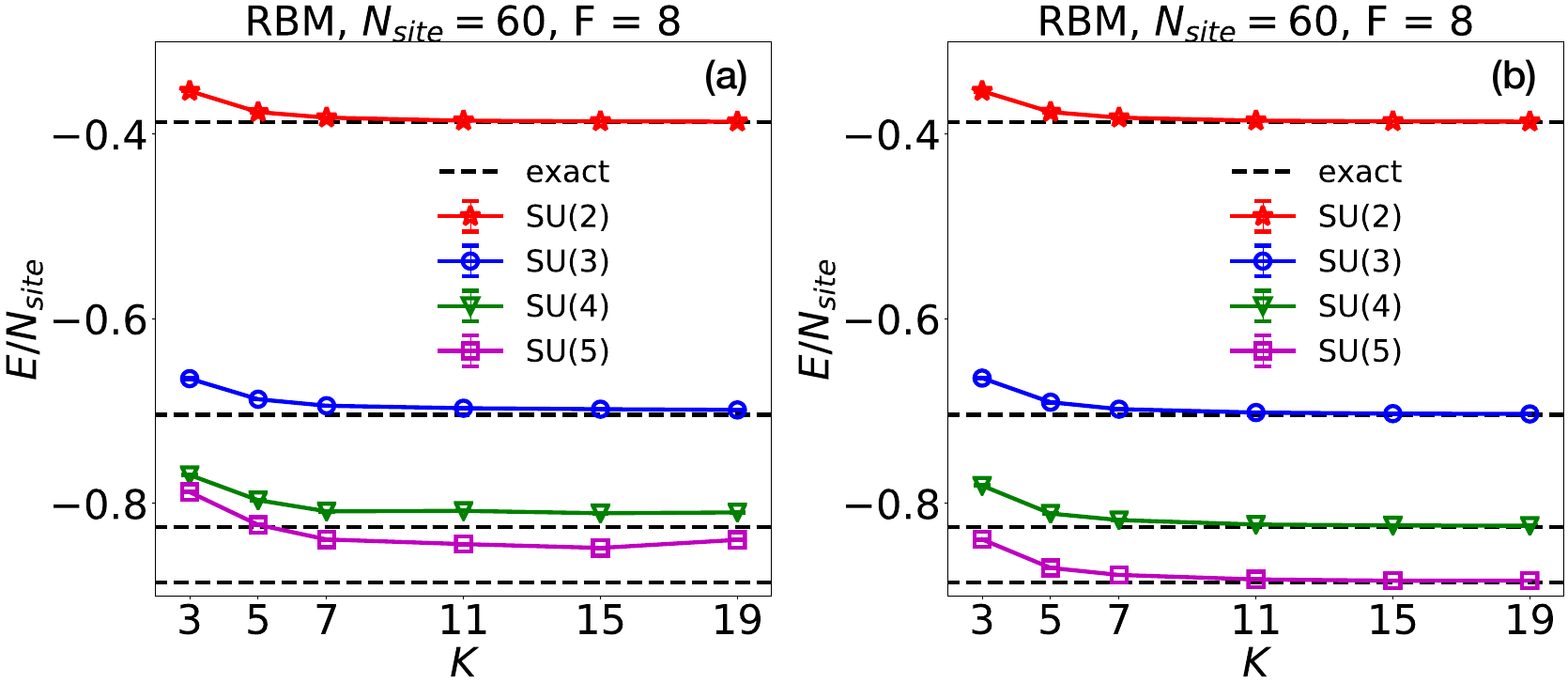}
\caption{The RBM ground state energies for 60 sites SU($N$) spin chains, where $N = 2, 3, 4, 5$. (a) is for state value encoding. (b) is for state one-hot encoding. The only variable parameter is the kernel size K. The black dashed lines are exact results from Bethe Ansatz.}\label{energyResultsRBM}
\end{figure}

So far, Restricted Boltzmann Machines (RBMs) remain as the most widely used machine learning tool in solving quantum many-body physics. In representing wave functions, an RBM can be regarded as a 1-hidden layer neural network \cite{Choo2018}, with activation function being $\ln(2\cosh(x))$. Also, the RBM wave function in Ref.~\cite{Carleo2017} uses translational symmetry, which is equivalent to a CNN with kernel size $K = N_\text{site}$ and periodic boundary condition. The number of channels $\alpha$ in Ref.~\cite{Carleo2017} plays the same role as the number of filters $F$ in our work. We plot the ground state variational energies of a convolutional RBM ($K \le N_\text{site}$) with different $K$ in Fig.~\ref{energyResultsRBM}, which should be compared with Fig.~\ref{energyResults}(a)(c) in the main text, where we plot the energies as a function of convolutional kernal size $K$ in a one-layer CNN. It can be seen that the RBM results are very similar to those of the one-layer CNN although they use different activation functions. In addition, Fig.~\ref{energyResultsRBM} also shows that, just as in our work, one-hot encoding with RBM has significant advantage over value encoding for $N > 2$.

\section{Comparison of GPU and CPU performance}
To compare the computational speed of the ISGO method on GPUs and CPUs, the wall time for the calculations of the SU(2) model on a $N_\text{site}=60$ size cluster with model architecture (L, F, K) = (16, 8, 3) is plotted in Fig.~\ref{offPolicy}(c) in the main text. We have performed simulations on a single K80 GPU with 11GB memory of Google Colaboratory and two 2.30GHz Intel(R) Xeon(R) CPUs. The ISGO method has two steps: Monte Carlo sampling and importance sampling gradient optimization. The Monte Carlo sampling step is always calculated on CPUs. The simulation with $N_\text{optimize} = 1$ corresponds to conventional GO method, whose performance on GPU (blue line) is similar to the one of the ISGO method on CPUs (green line), as in main text Fig.~\ref{offPolicy}(c). Remarkably, the simulation with the ISGO method on GPUs (red line) arrives at a good variational energy at about 0.5h, while the conventional GO method (blue line) requires about 5h to obtain the similar accuracy for variational energy. The simulation with the conventional GO method on CPUs (magenta line) is the slowest one due to python overhead. The results shows that the simulations with the ISGO method on GPU are at least one order of magnitude faster than the conventional GO method on any hardware. We also observe the speedup is even greater for more complex networks.

\end{document}